\documentclass[preprint2]{aastex}
\usepackage{natbib}
\usepackage{soul}

\shorttitle{Structure of SMGs}
\shortauthors{Iono et al.}

\begin{document}

\title{Clumpy and Extended Starbursts in the Brightest Unlensed Submillimeter Galaxies}

\author{Daisuke Iono\altaffilmark{1,2}, 
Min S. Yun\altaffilmark{3},
Itziar Aretxaga\altaffilmark{4},
Bunyo Hatsukade\altaffilmark{1},
David Hughes\altaffilmark{4},
Soh Ikarashi\altaffilmark{5},
Takuma Izumi\altaffilmark{6},
Ryohei Kawabe\altaffilmark{1},
Kotaro Kohno\altaffilmark{6,7},
Minju Lee\altaffilmark{8},
Yuichi Matsuda\altaffilmark{1,2},
Kouichiro Nakanishi\altaffilmark{1,2},
Toshiki Saito\altaffilmark{8},
Yoichi Tamura\altaffilmark{6},
Junko Ueda\altaffilmark{9},
Hideki Umehata\altaffilmark{6},
Grant Wilson\altaffilmark{3},
Tomonari Michiyama\altaffilmark{2},
Misaki Ando\altaffilmark{2}
}

\altaffiltext{1}{National Astronomical Observatory of Japan, National Institutes of Natural Sciences, 2-21-1 Osawa, Mitaka, Tokyo 181-8588}
\altaffiltext{2}{SOKENDAI (The Graduate University for Advanced Studies), 2-21-1 Osawa, Mitaka, Tokyo 181-8588}
\altaffiltext{3}{University of Massachusetts, Department of Astronomy, 710 North Pleasant Street Amherst, MA 01003}
\altaffiltext{4}{Instituto Nacional de Astrofisica, Optica y Electronica (INAOE), Luis Enrique Erro 1, Sta. Ma. Tonantzintla, Puebla, Mexico}
\altaffiltext{5}{Kapteyn Astronomical Institute, University of Groningen, PO Box 800, 9700AV Groningen, the Netherlands}
\altaffiltext{6}{Institute of Astronomy, The University of Tokyo, 2-21-1 Osawa, Mitaka, Tokyo 181-0015, Japan} \altaffiltext{7}{Research Center for the Early Universe, The University of Tokyo, 7-3-1 Hongo, Bunkyo, Tokyo 113-0033, Japan}
\altaffiltext{8}{Department of Astronomy, The University of Tokyo, 7-3-1 Hongo, Bunkyo-ku, Tokyo 133-0033, Japan}
\altaffiltext{9}{Harvard-Smithsonian Center for Astrophysics, 60 Garden Street, Cambridge, MA 02138, USA}

\begin{abstract}
The central structure in three of the brightest unlensed $z=3-4$ submillimeter 
galaxies are investigated through $0\farcs015$ -- $0\farcs05$ (120 -- 360~pc) 860$\micron$ continuum images 
obtained using the Atacama Large Millimeter/submillimeter 
Array (ALMA).  The distribution in the central kpc in AzTEC1 and AzTEC8 
are extremely complex, and they are composed of multiple $\sim 200$~pc clumps. 
AzTEC4 consists of two sources that are separated by $\sim 1.5$~kpc, indicating a mid-stage merger.
The peak star formation rate densities in the central clumps are 
$\sim300 - 3000$~M$_\odot$~yr$^{-1}$~kpc$^{-2}$, suggesting regions with 
extreme star formation near the Eddington Limit.
By comparing the flux obtained by ALMA and Submillimeter Array (SMA), 
we find that $68-90\%$ of the emission is 
extended ($\gtrsim$ 1~kpc) in AzTEC 4 and 8.
For AzTEC1, we identify at least 11 additional compact ($\sim 200$~pc) 
clumps  in the extended 3 -- 4~kpc region. 
Overall, the data presented here suggest that the luminosity surface densities observed at 
$\lesssim 150$~pc scales are  roughly similar to that observed in local ULIRGs, as in the 
eastern nucleus of Arp~220. 
Between 10 to 30\% of the 860$\micron$ continuum is concentrated in clumpy structures in the 
central kpc while the remaining flux is distributed over  $\gtrsim1$ kpc regions, some of which could also be 
clumpy.  These sources can be explained by a rapid inflow of gas such as a merger of 
 gas-rich galaxies, surrounded by extended and clumpy starbursts.  However, 
 the cold mode accretion model is not ruled out.

\end{abstract}


\section{Introduction}

Dusty star forming galaxies (or submillimeter galaxies; SMGs) represent a population 
of the most massive young galaxies rapidly 
building up their mass in the early universe \citep{2014PhR...541...45C}. 
The number density of the brightest SMGs with apparent  
submm flux in excess of 10 mJy at $850\micron$ is $\sim 5$~deg$^{-2}$
\citep{2016PASJ...68...36H}, and 
the intrinsic Star Formation Rate (SFR) can exceed 1000~M$_\odot$~yr$^{-1}$ 
if the submillimeter flux arises entirely from star formation.  
Although they are rare, these SMGs host the most intense starbursts in the entire universe and may  
represent the formation sites of massive elliptical galaxies.

A merger of two gas-rich, massive spirals can account for the origin of the 
starburst galaxies with SFR $\sim100$~M$_{\odot}$~yr$^{-1}$.
However, SMGs with SFR in excess of $1000$~M$_\odot$~yr$^{-1}$ are 
difficult to explain with the conventional major merger scenario alone.
Whether they are fueled and sustained through continuous accretion of cold gas \citep[e.g.][]{2005MNRAS.363....2K} 
and forming stars in wide-spread clumpy gas-rich disks \citep[e.g.][]{2014ApJ...780...57B},  through
gas-rich major mergers \citep[e.g.][]{2012MNRAS.424..951H,2015A&A...584A..32M}
or through continuous infall of gas ejected via  
stellar feedback \citep{2015Natur.525..496N},  
understanding the physical origin of the intense star formation activity  
can only be achieved by high resolution observations 
of the dust and gas.

The size scale and the spatial distribution of the submm continuum 
in the bright SMGs are recently being studied  
using the Atacama Large Millimeter/submillimeter Array (ALMA). 
The sizes of the SMGs with L$_{\rm IR} \sim 10^{12-13}$~L$_{\odot}$
have been constrained to 1.6 - 2.4~kpc 
from $0\farcs1 - 0\farcs7$ observations \citep{2015ApJ...810..133I, 2015ApJ...799...81S,2016arXiv160107549O}, 
and the compactness suggests that they  
may become the compact quiescent galaxies found at $z\sim2$ \citep{2014ApJ...782...68T}.
The extended baseline capabilities of ALMA are now 
providing images at $< 0\farcs05$ resolution \citep{2015ApJ...808L...4A}, 
 allowing us to directly probe the central kpc of these SMGs.
Here we present $0\farcs015 - 0\farcs05$ resolution ALMA images in  
three of the brightest unlensed $z=3 - 4$ SMGs in the COSMOS field -- 
AzTEC1, AzTEC4 and AzTEC8.  These sources were previously confirmed to be 
compact from $0\farcs5$ images obtained at the 
Submillimeter Array \citep[SMA;][]{2004ApJ...616L...1H,2008ApJ...688...59Y,2010MNRAS.407.1268Y}  
with no significant evidence of lensing by a foreground galaxy.
All three sources were first discovered in the COSMOS field \citep{2007ApJS..172...38S}
using the AzTEC camera mounted on the James Clerk Maxwell Telescope (JCMT) 
\citep{2008MNRAS.385.2225S}.  

AzTEC1 has a SFR of $1300$~M$_{\odot}$~yr$^{-1}$, 
 dust mass of $(3.7 \pm 0.7) \times 10^{11}$~M$_{\odot}$ assuming $T_d = 35$~K, 
 and a stellar mass of $\sim 10^{11}$~M$_{\odot}$ \citep{2011ApJ...731L..27S, 2015MNRAS.454.3485Y}.
The redshift of AzTEC1 is spectroscopically confirmed at 
 $z = 4.3420 \pm 0.0004$  \citep{2015MNRAS.454.3485Y}.
  AzTEC 4 has  a tentative hard X-ray source  \citep{2010MNRAS.407.1268Y}, 
 and a $B$-band/$i$-band counterpart located $\sim0\farcs5$ away. 
The photometric redshift is $z  \sim 4$ \citep{2011ApJ...731L..27S,2014ApJ...782...68T}.
A fit to the SED yields a  SFR of $1778$~M$_{\odot}$~yr$^{-1}$,
 dust mass of $4.0 \times 10^{9}$~M$_{\odot}$, 
 IR luminosity of $1.7 \times 10^{13}$~L$_{\odot}$, 
 and a stellar mass of $1.6 \times 10^{11}$~M$_\odot$ \citep{2014ApJ...782...68T}.
 In contrast to AzTEC4, 
 AzTEC8 does not have a significant X-ray 
counterpart, but has a bright and extended VLA 20-cm emission to the east. 
The spectroscopic redshift of this source is $z = 3.179$ 
\citep{2012A&A...548A...4S} while the nearby radio source has a CO redshift of 
 $z = 1.950$ (Yun et al., in prep).
A fit to the SED yields a SFR of $2818$~M$_{\odot}$~yr$^{-1}$,
 dust mass of $5.0 \times 10^{9}$~M$_{\odot}$, 
  IR luminosity of $2.8 \times 10^{13}$~L$_{\odot}$, 
 and a stellar mass of $3.2 \times 10^{11}$~M$_\odot$ \citep{2014ApJ...782...68T}.

We adopt the WMAP concordance $\Lambda$CDM cosmology parameters with 
H$_0 = 69.6$~km~s$^{-1}$~Mpc$^{-1}$ and  $\Omega_M = 0.286$ \citep{2014ApJ...794..135B}.

\section{ALMA Observations}

Observations were carried out on November 6 -- 8, 2015 
using 46 - 47 antennas with projected baselines that ranged from 100~m to 14~km.
The delivered manually calibrated  data product 
were  inspected in the visibilities and in the dirty image, 
and we flagged problematic antennas that were causing 
ripples in the image.
 We used CASA v4.5.1 for imaging.
We placed CLEAN boxes to the regions where clear ($\geq 3\sigma$) emission 
features are seen, and  CLEANed down to 1.5$\sigma$ in order to properly 
image the extended emission.
We searched for the optimal beam size by 
varying the Robust weighting.  
We tapered the AzTEC4 data to $0\farcs05$ since the emission is only 
marginally detected with the native resolution. 
The continuum image was generated by 
using all of the bands except for AzTEC1 in which the frequency range of the 
redshifted [\ion{C}{2}]  emission was excluded. 
For AzTEC1 and AzTEC8, 
we searched for the [\ion{C}{2}] and 
[\ion{N}{2}] emission by constructing  
images with 100 km~s$^{-1}$  resolution with a $0\farcs06$ beam.  
Both lines were undetected with $3\sigma$ upper limits of 1.4~mJy for AzTEC1 and 1.0~mJy for AzTEC8.
The center frequency 
of the four correlator basebands was 350.739~GHz (855.337~$\mu$m).

\begin{figure*}  [t]
	\begin{center}
		\includegraphics[scale=0.85]{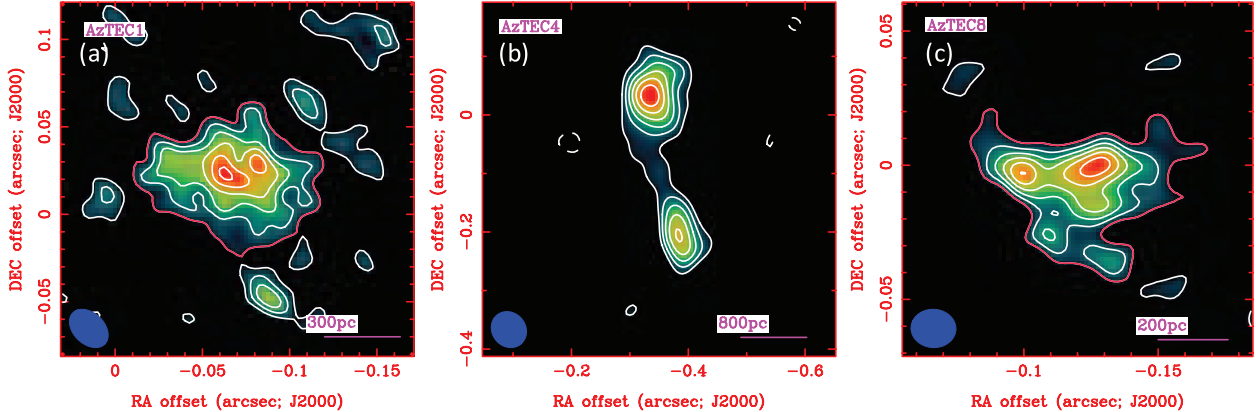}
        		\caption{860$\mu$m images of (a) AzTEC1 (Robust=2 and no tapering), (b) AzTEC4 (Robust=2 and \textit{outertaper}=$0\farcs05$),  and (c) AzTEC8 (Robust=0.5).  The physical scale shown in the lower right corner of each panel assumes $z=4.342$ for AzTEC1, $z=4$ for AzTEC4 and $z=3.179$ for AzTEC8.  The lowest contours represent $3\sigma$, and they increase in steps of $1\sigma$ (the 1$\sigma$ values are shown in Table~\ref{tab1}). Negative contours are shown in dashed lines. The total flux enclosed within the red $3\sigma$ contour for AzTEC1 and 8 are used to derive the parameters in Table~\ref{tab2}.  All of the images are referenced from the SMA coordinates.}
	        \label{fig1}
	\end{center}
\end{figure*}

\begin{figure*}  [t]
	\begin{center} 
		\includegraphics[scale=0.8] {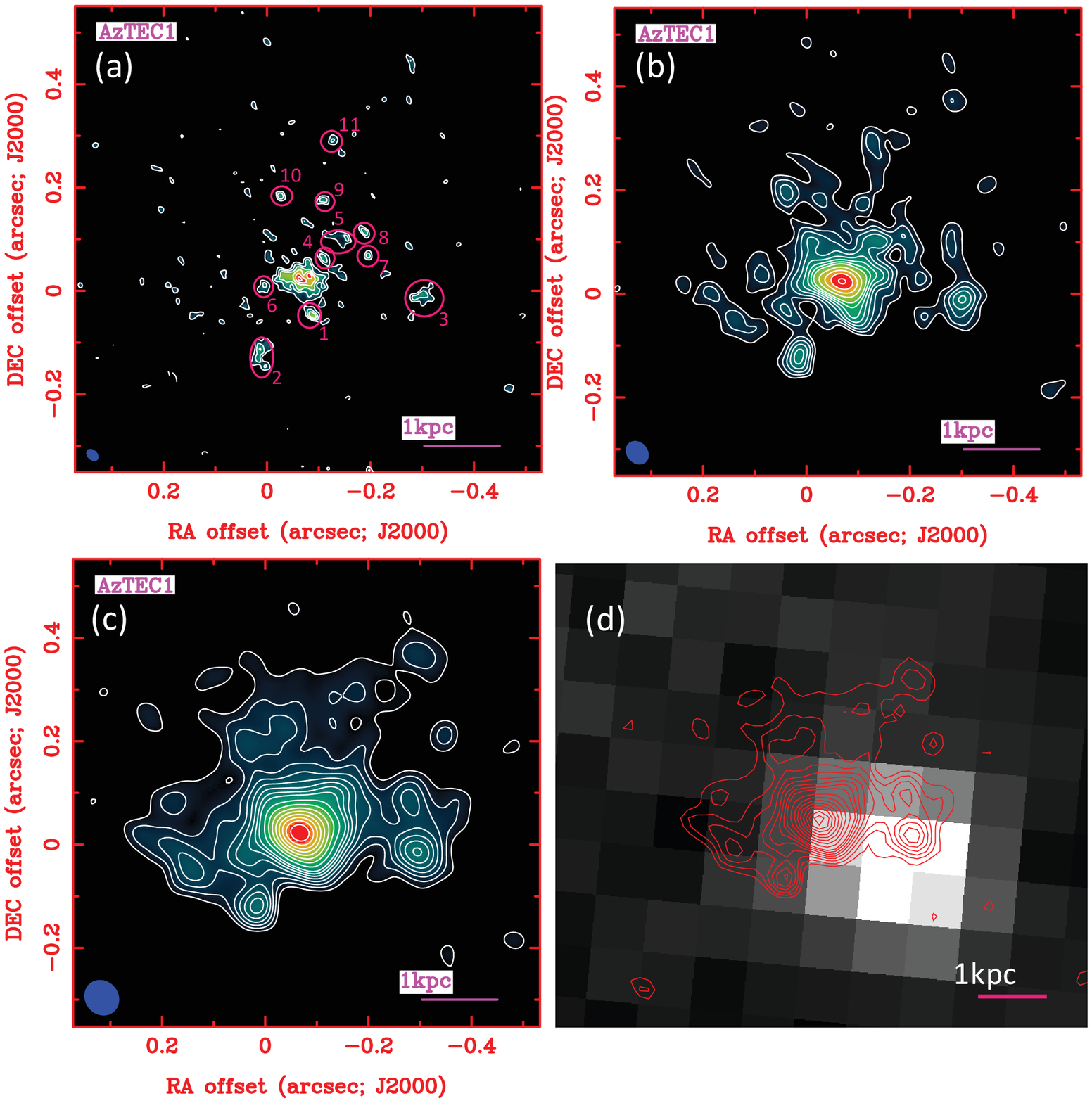}		
        		\caption{860$\mu$m images of AzTEC1 with (a) $0\farcs026 \times 0\farcs018$, (b) $0\farcs048\times 0\farcs039$,  and (c) $0\farcs070 \times 0\farcs063$ resolution.  The lowest contours represent $3\sigma$, and they increase in steps of $1\sigma$ up to $10\sigma$ and steps of $2\sigma$ beyond.  The properties of the clumps labeled 1 to 11 in (a) are presented in Table~\ref{tab3}. 
		(d) Overlay of ALMA contours on the HST F160W image.  The contour levels are the same as in (c).  We note a $\sim 0\farcs2$ offset between the ALMA peak and the HST peak \citep[see also][]{2016arXiv160605351O}.  Detailed investigation of the relative astrometry is differed to a future analysis.}
	        \label{fig2}
	\end{center}
\end{figure*}

An astrometry check source (J0930+0034)
was observed several times within the main 
observing sequence.
The emission peak of J0930+0034 in all datasets were found within
$0\farcs0015 - 0\farcs0045$ of the phase center, and we adopt this value as the 
astrometric accuracy.
We note that 
the peaks of all three sources were detected $\sim 0\farcs05$ to $0\farcs4$ west of the 
phase center defined from the SMA position.   The cause of this offset is unknown at present.

For AzTEC1, 
we obtained the $0\farcs3$ resolution data from the ALMA Archive 
(2012.1.00978.S) and combined it with our high resolution data.  The $0\farcs3$ resolution
image alone is marginally resolved, with a total flux of 15.4~mJy.  
The center frequency was 343.769~GHz (872.679~$\mu$m), 
which is 7~GHz lower than the adopted frequency for the high resolution data.
Therefore, 
it is not tuned to the frequency of the
redshifted [\ion{C}{2}] line.
Data combination was performed after down-weighting the $0\farcs3$ resolution data
by 1/10, yielding a synthesized beam with reasonable sidelobe levels.  The combined visibilities were 
then imaged using the multiscale CLEAN option in CASA, first without any visibility tapering, and then by 
varying the \textit{outertaper} parameter from $0\farcs03 - 0\farcs05$. 
The observed properties are presented in Table~\ref{tab1}.

\begin{deluxetable}{ccccccccccc}
\tabletypesize{\scriptsize}
\tablecaption{Observational Properties \label{tab1}}
\tablewidth{0pt}
\tablehead{
\colhead{Source} & \colhead{redshift} & 1 arcsec & \colhead{RA} & \colhead{DEC} & \colhead{RMS} & \colhead{beam (Position Angle)}  & \colhead{Robust\tablenotemark{a}} &\colhead{$S_{860}$\tablenotemark{b}}  \\
& & [kpc] & \colhead{(J2000)} & \colhead{(J2000)} & \colhead{[$\mu$Jy]} & & &\colhead{[mJy]} 
}
\startdata
AzTEC1 & 4.342 & 6.7 & 9:59:42.86  &  +2:29:38.2 & 31    &  $0\farcs026 \times 0\farcs018$ (46$^\circ$)  & 2 & $3.67 \pm 0.37$ \\
 & &  &   &   &  38 & $0\farcs048 \times 0\farcs039$ (42$^\circ$) & 2 ($0\farcs03$) &    $9.09 \pm 0.91$ &  \\
 & &  &   &   &  56 & $0\farcs070 \times 0\farcs063$ (41$^\circ$) & 2 ($0\farcs05$) &    $14.16 \pm 1.42$ &  \\
AzTEC4  & $\sim 4$\tablenotemark{c}  & 7.1 & 9:59:31.72  &  +2:30:44.0 &  61    &   $0\farcs064 \times 0\farcs057$  (40$^\circ$) & 2 ($0\farcs05$) &  $3.26 \pm 0.40$ \\
AzTEC8  & 3.179   & 7.7 & 9:59:59.34  &  +2:34:41.0 &  34   &   $0\farcs017 \times 0\farcs014$ (87$^\circ$) & 0.5 &  $1.63 \pm 0.16$\\
\enddata
\tablenotetext{a}{The value of Robust parameter used in CASA.  The values in the () are used for the \textit{outertaper} in the CASA task CLEAN.}
\tablenotetext{b}{Total ALMA flux above $3\sigma$. For AzTEC1, we integrate the $\geq 3\sigma$ emission in Figure~2. The SMA 880 micron fluxes (beam sizes) are $13.8 \pm 2.3$ ($0\farcs86 \times 0\farcs55$),  $13.1\pm1.8$ ($0\farcs86 \times 0\farcs77$), and $17.7 \pm 2.3$~mJy ($0\farcs86 \times 0\farcs55$), for AzTEC1, AzTEC4, and AzTEC8, respectively  \citep{2008ApJ...688...59Y,2010MNRAS.407.1268Y}.}
\tablenotetext{c}{The photometric redshifts derived in the literature are $4.70^{+0.43}_{-1.11}$ \citep{2007ApJ...671.1531Y} and $4.93^{+0.43}_{-1.11}$ \citep{2012A&A...548A...4S,2014ApJ...782...68T}.  We adopt $z=4$ for all calculations in this paper.}
\end{deluxetable}

\section{Characteristics of the Individual Sources}

The ALMA images and the derived parameters are presented in Figure~\ref{fig1} -- \ref{fig2}, and Table~\ref{tab2}.
We derive the SFR of each component by scaling the total SFR with the ratio between the derived 
flux densities and the total SMA flux.  
The luminosity ($\Sigma_{\rm L_{FIR}}$) and SFR surface densities ($\Sigma_{\rm SFR}$) 
are derived by using the observed source size measurements presented in Table~\ref{tab2}.
The peak brightness temperatures using the optically thick
Planck equation are 33 - 82~K. 

\begin{deluxetable}{llcccccccc}
\tabletypesize{\scriptsize}
\tablecaption{Star Formation Properties in the Central kpc\label{tab2}}
\tablewidth{0pt}
\tablehead{
\colhead{Source} & \colhead{Component} & \colhead{$S_{860}$}& \colhead{Size} & \colhead{L$_{\rm FIR}$} & 
\colhead{$\Sigma_{L_{\rm FIR}}$} & \colhead{SFR} & \colhead{$\Sigma_{\rm SFR}$} \\
 &  & \colhead{[mJy]} & \colhead{[kpc$^2$]} & \colhead{[L$_\odot$]} &   \colhead{[L$_\odot$~kpc$^{-2}$]} 
 & \colhead{[M$_\odot$~yr$^{-1}$]} & \colhead{[M$_\odot$~yr$^{-1}$~kpc$^{-2}$]}  \\
}
\startdata
AzTEC1 & Total\tablenotemark{a}& $14.16\pm1.42$ & $10.15\pm1.62$ & $1.1 \times 10^{13}$ &  $(1.1 \pm0.2) \times 10^{12}$ & 1300 & $130 \pm 30$  \\
 & Integrated\tablenotemark{b} & $1.54\pm0.15$ & $0.27\pm0.04$ & $(1.2 \pm 0.2) \times 10^{12}$ & $(4.6 \pm 3.0) \times 10^{12}$& $150 \pm 40$&  $540 \pm 160$ \\
& Central Peak & $0.25 \pm 0.03$ & - & $(2.0 \pm 0.4) \times 10^{11}$ & $(8.1 \pm 1.7) \times 10^{12}$& $23 \pm 6$ & $960 \pm 240$ \\
& Western Peak & $0.23\pm0.03$ & - & $(1.8 \pm 0.4) \times 10^{11}$ & $(7.4 \pm 1.6) \times 10^{12}$& $21 \pm 5$ & $870 \pm 220$ \\
AzTEC4\tablenotemark{c} & Total\tablenotemark{d}  & $13.1\pm1.8$ & - & $1.7 \times 10^{13}$ & -- & $1778 \pm 733$ & -- \\
& Northern Source & $1.98 \pm 0.31$ & $0.64 \pm 0.17$ & $ (2.6 \pm 0.5) \times 10^{12}$ & $(4.0 \pm 1.9) \times 10^{12}$ & $270 \pm 120$ & $420 \pm 220$ \\
& Southern Source & $1.28\pm0.25$ & $0.42 \pm 0.17$ & $(1.7 \pm 0.4) \times 10^{12}$ & $(3.7 \pm 3.5) \times 10^{12}$ & $170 \pm 80$ & $410 \pm 250$ \\
& North+South & $3.26 \pm 0.40$ &  $1.06 \pm 0.24$ & $(4.2 \pm 0.8) \times 10^{12}$ & $(4.0 \pm 1.1) \times 10^{12}$ & $440 \pm 200$ & $420 \pm 210$ \\
& Northern Peak & $0.54 \pm 0.06$ & - & $(7.0 \pm 1.2) \times 10^{11}$ & $(3.3 \pm 0.6) \times 10^{12}$ & $73 \pm 33$ & $350 \pm 160$ \\
& Southern Peak & $0.44 \pm 0.06$ & - & $(5.6 \pm 1.1) \times 10^{11}$ & $(2.7 \pm 0.5) \times 10^{12}$  & $59 \pm 27$ & $280 \pm 130$ \\
AzTEC8\tablenotemark{c} & Total\tablenotemark{d}  & $17.7 \pm 2.3$ & - &$2.8 \times 10^{13}$ & -- & $2818 \pm 66$ & -- \\
& Integrated\tablenotemark{b}   & $1.63 \pm 0.16$ & $0.16\pm0.02$ & $(2.6 \pm 0.4) \times 10^{12}$ & $(1.6 \pm 1.3) \times 10^{13}$ & $260 \pm 40$ & $1620 \pm 330$ \\
& Western Peak & $0.30 \pm 0.03$ & - & $(4.7 \pm 0.6) \times 10^{11}$ & $(2.9 \pm 0.4) \times 10^{13}$ & $47 \pm 8$ & $2950 \pm 520$ \\
& Eastern Peak & $0.27 \pm 0.03$ & - & $(4.3 \pm 0.6) \times 10^{11}$ & $(2.7 \pm 0.4) \times 10^{13}$ & $43 \pm 8$ & $2720 \pm 490$ \\
\enddata
\tablenotetext{a}{Using the size of the region enclosed within the $3\sigma$ contours in Figure~\ref{fig2}~(c).  The IR luminosity and SFR are obtained from \citet{2015MNRAS.454.3485Y}.}
\tablenotetext{b}{Using the total ALMA flux in the region enclosed within the red $3\sigma$ contour in Figure~\ref{fig1}.  }
\tablenotetext{c}{The IR luminosity and SFR are obtained from \citet{2014ApJ...782...68T}.}
\tablenotetext{d}{Using the SMA flux.}

\end{deluxetable}

\textit{AzTEC1--}
From the $0\farcs026 \times 0\farcs018$ resolution image (Figure~\ref{fig1}~(a)), we find that 
the central $\lesssim 0.5$~kpc region of AzTEC1 is characterized by two central peaks, each with 
$249\pm31 \mu$Jy~beam$^{-1}$ and $227\pm31\mu$Jy~beam$^{-1}$ and a separation of $\sim150$~pc.   
The double peak structure is surrounded by a complex emission that extends to $\lesssim 0.5$~kpc scale, 
and the total flux enclosed within the red $3\sigma$ contours is $1.54 \pm 0.15$~mJy.  

The $>$~0.5~kpc region of AzTEC1 is characterized by two 
distinct features; (1)  smooth and extended 
emission that reaches 3 -- 4~kpc to the north (see Figure~\ref{fig2}~(c)), 
and (2) numerous $\sim 200$~pc scale clumps 
found in the extended region  (see Figure~\ref{fig2}~(a)). 
The excellent consistency between the total flux 
in the $0\farcs070 \times 0\farcs063$ image 
in Figure~\ref{fig2}~(c) ($14.16 \pm 1.42$~mJy)
and the SMA image ($13.8 \pm 2.3$~mJy) suggests that 
much of the extended features are real. 
In order to quantify the size of the emission in the central kpc and the associated flux, 
the peak emission of Figure~\ref{fig2}~(c)  
was modeled with a two dimensional Gaussian. We obtain $(0\farcs102 \pm 0\farcs007) \times (0\farcs083 \pm 0\farcs006)$ which is equivalent to $(700 \pm 50) \times (570 \pm 40)$~pc, with an integrated flux density of $4.52 \pm 0.23$~mJy (32\% of the total flux).  This suggests that the extended structure contains 68\% ($9.64 \pm 1.44$~mJy) of the total flux.  The half light radius is $1.1 \pm 0.1$~kpc.

In the $>$~0.5~kpc region, we have manually identified $\sim40$ compact clumps 
which are detected at $3\sigma$ or above. 
The number reduces to 11 clumps if we set the threshold to $4\sigma$.
The number of negative $3\sigma$ and $4\sigma$ clumps in the same region are two and one, respectively.
The sources sizes, integrated flux densities, SFR, 
and the  $\Sigma_{\rm SFR}$ are summarized in Table~\ref{tab3}.
The average source size is $(0.30 \pm 0.05) \times (0.17 \pm 0.03)$~kpc, and the 
total flux from all of  the $\geq 4\sigma$ clumps is $2.69 \pm 0.27$~mJy, %
which translates to a SFR of $250 \pm 30$~M$_\odot$~yr$^{-1}$.

\textit{AzTEC4} -- 
At $0\farcs05$ resolution, the emission in AzTEC4 is divided into two distinct clumps 
with a separation of $\sim1.5$~kpc in the north-south direction. 
The large separation between the two clumps suggest that AzTEC4 is a mid stage merger.
The peaks of the northern and southern sources are $536\pm61\mu$Jy~beam$^{-1}$ 
and $435\pm61\mu$Jy~beam$^{-1}$, and the    
deconvolved source sizes from a two dimensional Gaussian fit are 
$(0\farcs130 \pm 0\farcs023) \times (0\farcs086 \pm 0\farcs017)$ 
 and $(0\farcs132 \pm 0\farcs030) \times (0\farcs056 \pm 0\farcs018)$, respectively.
We have further inspected the residuals after subtracting both sources 
from the image, and found that the image is dominated by noise.
The total integrated flux of the northern and 
southern sources are $1.98 \pm 0.31$~mJy and $1.28 \pm 0.25$~mJy, respectively, 
yielding a total of $3.26 \pm 0.40$~mJy.  
The total flux derived from the $0\farcs86 \times 0\farcs77$ SMA beam is 
$13.1 \pm 1.8$~mJy for a Gaussian model \citep{2010MNRAS.407.1268Y}.
Therefore, $75\%$ of the total flux is missing in the extended structure. 

 \textit{AzTEC8} -- 
 The central region is clearly resolved into two clumps separated by $\sim 200$~pc in the
 east-west direction.  
 The peak of the eastern and western clumps are
$273\pm34\mu$Jy~beam$^{-1}$ and $296\pm34\mu$Jy~beam$^{-1}$, respectively.  
The total integrated flux 
in the central kpc (region within the red contour in Fig~\ref{fig1}) is $1.63\pm0.16$~mJy.  
The total flux derived from the $0\farcs86 \times 0\farcs55$ SMA beam is 
$17.7 \pm 2.3$~mJy for a Gaussian model \citep{2010MNRAS.407.1268Y}, and thus 
 $\sim 90\%$ of the flux is in the extended structure.

\section{Nature of the Central Star Forming Region}
\subsection{Eddington Limited Starbursts?}

The stellar radiation pressure imposed upon the surrounding optically thick dust and gas    
can provide an important stabilization mechanism for gas against self gravity.
Assuming a  one-zone moderately optically thick disk model, a Toomre-stable (Q $\sim 1$) 
disk,  and  T $<100$~K, 
the $\Sigma_{\rm SFR}$ required to support the disk with radiation pressure is
$\sim 1000$~M$_{\odot}$~yr$^{-1}$~kpc$^{-2}$ \citep{2005ApJ...630..167T}.  At
higher temperatures (T = 100 - 200~K), which may be more appropriate at the centers   
of ULIRGs \citep{2014ApJ...789L..36W} and SMGs, the correlation between opacity and temperature 
breaks down, 
and $\Sigma_{\rm SFR}$ becomes a function of various model parameters such as 
rotation velocity, gas mass fraction and opacity.  
We lack sufficient data, such as gas kinematics, to constrain these parameters properly, 
and the required $\Sigma_{\rm SFR}$ can be 
$\sim 100 - 2000$~M$_{\odot}$~yr$^{-1}$~kpc$^{-2}$ \citep[see Equation 28 of][]{2005ApJ...630..167T}
even  with a conservative set of assumptions.

The integrated and beam averaged peak  $\Sigma_{\rm SFR}$ range from 
$280 \pm 130$ to $960 \pm 240$~M$_\odot$~yr$^{-1}$~kpc$^{-2}$ 
for AzTEC1 and AzTEC4.
On the other hand, the $\Sigma_{\rm SFR}$ of the entire AzTEC1
is much lower ($130 \pm 30$~M$_\odot$~yr$^{-1}$~kpc$^{-2}$)
and it is close to the lower end of the 
predicted $\Sigma_{\rm SFR}$ for a radiation pressure supported disk.  
 These are consistent with the average $\Sigma_{\rm SFR}$ found in the   
most luminous starburst galaxies observed in the early universe 
(80 - 1000 M$_\odot$~yr$^{-1}$~kpc$^{-2}$) \citep{2006ApJ...640..228T,2009Natur.457..699W,2013Natur.496..329R,
2014ApJ...796...84R,2015ApJ...798L..18H,2016arXiv160107549O}.
The gravitationally lensed source SDP.81  
(SFR $\sim 500$~M$_\odot$~yr$^{-1}$) was observed at 200~pc resolution  \citep{2015ApJ...808L...4A}, 
and the $\Sigma_{\rm SFR}$ of the 14 clumps range from $\sim$100 to 400~M$_{\odot}$~yr$^{-1}$~kpc$^{-2}$ 
\citep{2015PASJ...67...93H}, which is slightly lower than the central regions of AzTEC1 and AzTEC8.
In contrast to AzTEC1 and AzTEC4, the average $\Sigma_{\rm SFR}$ in AzTEC8 is 
close to the upper end of the model predictions.  The beam averaged peak values 
are even higher ($\sim 3000$~M$_\odot$~yr$^{-1}$~kpc$^{-2}$).  
The extreme values derived in AzTEC8 may suggest
that it is forming stars near or above the Eddington limit.

\subsection{Comparison with local U/LIRGs}

The nearby late stage merging ULIRG Arp~220 was 
recently observed using ALMA at 
$0\farcs36 \times 0\farcs20$ \citep{2014ApJ...789L..36W}
and $0\farcs32 \times 0\farcs28$ \citep{2015ApJ...800...70S} resolution at 690GHz.  
Assuming $1\arcsec$ = 475~pc, these correspond
to 100 -- 170~pc scales, allowing us a direct comparison with the 
physical properties of AzTEC1 and AzTEC8. 
The luminosity surface densities in Arp 220 are 
$\geq 10^{14}$~L$_\odot$~kpc$^{-2}$ and $10^{12-13}$~L$_\odot$~kpc$^{-2}$ 
for the western and eastern nucleus, 
respectively \citep{2014ApJ...789L..36W} 
\citep[see also][]{2007A&A...468L..57D,2008ApJ...684..957S,2015ApJ...799...10B}.
The average luminosity 
densities derived in the central peaks of AzTEC1 [($7.8 \pm 2.3) \times 10^{12}$~L$_\odot$~kpc$^{-2}$] 
and AzTEC8 [($2.8 \pm 0.6) \times 10^{13}$~L$_\odot$~kpc$^{-2}$] are consistent 
with the eastern nucleus, but an order of magnitude lower than
the western nucleus of Arp220.
The high value in the western nucleus may be attributed to the presence of an AGN. 
In contrast, the same quantities derived in the less luminous mid-stage merging LIRG  
VV114 is at least 2 -- 3 orders of magnitude lower \citep{2015ApJ...803...60S}.

\section{Discussion and Summary}
\subsection{Nature of the Clumpy and Extended Emission in AzTEC1}

The important new finding from the AzTEC1 map is the presence of compact ($\sim 200$~pc) 
clumps and the diffuse and extended emission out to 3-4~kpc.

\textit{Compact Clumps--} 
These clumps can grow to contain a stellar mass of $\sim 10^9$~M$_\odot$ by $z=2$~(2~Gyr), 
even if they are converting gas to stars at an efficiency of 10\% \citep{1982MNRAS.200..159L}.
Numerical simulations of isolated gas rich disks suggest that gas 
clumps of $\sim 10^9$~M$_\odot$ survive for at least $10^8$~years 
during the course of their lifetime \citep{2014ApJ...780...57B}.  
These massive clumps can migrate inwards via dynamical friction 
and coalesce with the central galaxy within a Gyr time scale
 \citep{1999ApJ...514...77N, 2012MNRAS.427..968H, 2012MNRAS.422.1902I}.
We note that 
the sizes and ubiquity of the clumps are similar to the young super star clusters 
found in the central region of Arp~220 \citep{1998ApJ...492L.107S,2001MNRAS.321...11S,2006ApJ...641..763W}.

\textit{Diffuse Extended Emission --} 
The extended emission accounts for 68\% (SFR~$= 880 \pm 130$~M$_\odot$~yr$^{-1}$)  
of the total flux.  
This is consistent with the finding by \citet{2015MNRAS.454.3485Y}  
whose SED modeling required  $\sim 50\%$ of the gas and dust to be outside the central kpc 
in order to account for the rest frame optical light.
The extended structure may suggest a presence of a $3-4$~kpc star-bursting disk, 
or a massive starburst driven outflow from the central kpc.
Alternatively, 
the extended material can quickly cool and settle into a compact star-forming disk.

\subsection{Emerging Picture}

The new ALMA data presented here suggests that the 
central region of 
some of the brightest 
unlensed SMGs in the universe (AzTEC1 and AzTEC8) 
are very compact ($\leq 0.5$~kpc) and contain at least two central clumps, 
forming stars near the Eddington limit, with 
physical properties that are similar to the eastern nucleus of Arp220.  
AzTEC4 consists of two sources that are separated by $\sim 1.5$~kpc, indicating a mid-stage merger.
In addition, we find that 68 -- 90\% of the emission is distributed over $\gtrsim1$ kpc regions in 
all three sources.

A major merger of two gas rich galaxies can explain the morphology and 
the physical properties seen in the central kpc of all three AzTEC sources.
However, this scenario alone cannot explain the large amount of extended emission.  
The extended structure with numerous clumps (particularly for AzTEC1) 
is reminiscent of an isolated clumpy star 
forming disk model which is often invoked to explain the morphology and
kinematics of high redshift star forming galaxies \citep[e.g.][]{2009ApJ...706.1364F,2014ApJ...780...57B}. 
However, the characteristic sizes of these disks 
are a factor of 2-3 larger than the sizes of the AzTEC sources ($\lesssim 3 - 4$~kpc). 
Although the sample size used in this study is small, we suggest 
a hybrid scenario, i.e. a merging central ULIRG surrounded by  clumpy and 
extended star-bursting material, to be the favorable model here.
The relative compactness of these sources supports the idea that 
these SMGs will become the 
compact massive galaxies 
found at $z\sim2$ \citep[e.g.][]{2012MNRAS.424..951H,2014ApJ...782...68T, 2015ApJ...810..133I, 2015ApJ...799...81S, 2015ApJ...813...23V,2015MNRAS.449..361W}.
Alternatively, the extended material can quickly reform a disk, becoming a bulge dominated 
disk galaxy \citep{2005ApJ...622L...9S,2009ApJ...691.1168H,2014ApJS..214....1U}.
Detailed kinematical information of gas will provide us with better insights to the overall picture.

\begin{deluxetable}{cccccccc}
\tabletypesize{\scriptsize}
\tablecaption{Sizes and Star Formation Properties of Clumps in AzTEC1\label{tab3}}
\tablewidth{0pt}
\tablehead{
\colhead{ID} &
\colhead{Source Size} &
\colhead{Source Size} &
\colhead{Flux Density} &
\colhead{SFR} &
\colhead{$\Sigma_{\rm SFR}$} \\
 & 
 \colhead{[arcsec]} & 
 \colhead{[kpc]} & 
  \colhead{[mJy]} & 
 \colhead{[M$_\odot$~yr$^{-1}$]} &
  \colhead{[M$_\odot$~yr$^{-1}$~kpc$^{-2}$]} 
}
\startdata
1 & $(0\farcs036 \pm 0\farcs008) \times (0\farcs021 \pm 0\farcs003)$ & $(0.25 \pm 0.06) \times (0.14 \pm 0.02)$  & $0.283 \pm 0.079$ & $27 \pm 9$ & $650 \pm 280$ \\
2 & $(0\farcs055 \pm 0\farcs014) \times (0\farcs028 \pm 0\farcs005)$ & $(0.38 \pm 0.09) \times (0.19 \pm 0.03)$  & $0.486 \pm 0.130$ & $46 \pm 14$ & $560 \pm 240$ \\
3 & $(0\farcs048 \pm 0\farcs013) \times (0\farcs029 \pm 0\farcs006)$ & $(0.33 \pm 0.09) \times (0.20 \pm 0.04)$  & $0.432 \pm 0.127$ & $41 \pm 14$ & $530 \pm 250$ \\
4 & $(0\farcs042 \pm 0\farcs012) \times (0\farcs021 \pm 0\farcs004)$ & $(0.29 \pm 0.08) \times (0.15 \pm 0.03)$  & $0.275 \pm 0.090$ & $26 \pm 10$ & $530 \pm 270$ \\
5 & $(0\farcs041 \pm 0\farcs013) \times (0\farcs023 \pm 0\farcs005)$ & $(0.28 \pm 0.09) \times (0.16 \pm 0.03)$  & $0.264 \pm 0.094$ & $25 \pm 10$ & $490 \pm 260$ \\
6 & $(0\farcs034 \pm 0\farcs010) \times (0\farcs028 \pm 0\farcs007)$ & $(0.23 \pm 0.07) \times (0.20 \pm 0.05)$  & $0.268 \pm 0.099$ & $25 \pm 10$ & $490 \pm 270$ \\
7 & $< 0\farcs026 \times < 0\farcs018$ & $< 0.18 \times < 0.12$ & $0.136 \pm 0.031$ & $13 \pm 4$ & $> 500$ \\
8 & $< 0\farcs026 \times < 0\farcs018$ & $< 0.18 \times < 0.12$ & $0.143 \pm 0.029$ & $13 \pm 4$ & $> 500$ \\
9 & $< 0\farcs026 \times < 0\farcs018$ & $< 0.18 \times < 0.12$ & $0.138 \pm 0.032$ & $13 \pm 4$ & $> 500$ \\
10 & $< 0\farcs026 \times < 0\farcs018$ & $< 0.18 \times < 0.12$ & $0.143 \pm 0.032$ & $13 \pm 4$ & $> 500$ \\
11 & $< 0\farcs026 \times < 0\farcs018$ & $< 0.18 \times < 0.12$ & $0.122 \pm 0.034$ & $11 \pm 4$ & $> 500$ \\
Mean\tablenotemark{a}  & $(0\farcs043 \pm 0\farcs008) \times (0\farcs025 \pm 0\farcs004)$ & $(0.30 \pm 0.05) \times (0.17 \pm 0.03)$  & $0.335 \pm 0.098$ & $32 \pm 9$ & $540 \pm 60$ \\
\enddata
\tablenotetext{a}{Mean and standard deviation of the sources that are resolved (i.e. ID=1-6)}.
\end{deluxetable}

\acknowledgments
We thank our referee for valuable comments that improved the contents of this paper significantly.
We  thank P. Cox, S. Iguchi, T. Kodama, D. Narayanan, J. Simpson for useful discussion. 
DI is supported by 
the 2015 Inamori Research Grants Program.
This work was performed in part at the Aspen Center for Physics, which is supported by National Science Foundation grant PHY-1066293.
This paper makes use of the following ALMA data:
   ADS/JAO.ALMA\#2012.1.00978.S and \#2015.1.01345.S. ALMA is a partnership of ESO (representing its member states),
   NSF (USA) and NINS (Japan), together with NRC (Canada), NSC and ASIAA (Taiwan), and KASI
   (Republic of Korea), in cooperation with the Republic of Chile.
   The Joint ALMA Observatory is operated by ESO, AUI/NRAO and NAOJ.


\end{document}